\documentclass{monog3}    % Comment the above and uncomment this
\usepackage{amssymb,graphicx}
\title{Problems of Condensed Matter Physics}

\author{\textit{Chapter author}: A. Author} %1     %%%%
\affiliation{A. Author's affiliation}

\author{A. Author2 and A. Author3}
\affiliation{Affiliation for Author2 and Author3}
\author{${}^{}$\\${}^{}$\\${}^{}$\\${}^{}$\\${}^{}$}  %2
\affiliation{${}^{}$\\${}^{}$\\${}^{}$\\${}^{}$\\${}^{}$}

\author{\textit{Editors}}  %3
\affiliation{}
\author{ Alexei L.\ Ivanov$^\dag$ and Sergei G.\ Tikhodeev$^\ddag$ }%4
\affiliation{\dag Department of Physics and Astronomy Cardiff UK\\
\ddag General Physics Institute Moscow Russia\\}
\authors{4}   %Number of author lines, please replace if you

\renewcommand{\thechapter}{~}
\renewcommand{\thesection}{\arabic{section}}

\renewcommand{\thefigure}{\arabic{figure}}

\begin{document}

%\maketitle

%\dedication{To Professor Leonid V.~Keldysh on his 75th anniversary}

%\input{Preface}

%\medskip

%\acknowledgements

%\tableofcontents

\maintext

%\documentclass[pre,twocolumn,
%nofootinbib, superscriptaddress,eqsecnum,
%amsfonts,amsmath,floatfix]{revtex4}

%\usepackage{epsfig}

%\begin{document}

%\title{
%Metal-insulator transition in a weakly interacting many-electron
%system with localized single-particle states for dummies
%}
%\author{D.~M.~Basko}
%\email{basko@phys.columbia.edu}
%\affiliation{Physics Department, Columbia University, New York, NY
%10027, USA}
%\author{I.~L.~Aleiner}
%\affiliation{Physics Department, Columbia University, New York, NY
%10027, USA}
%\author{B.~L.~Altshuler}
%\affiliation{Physics Department, Columbia University, New York, NY
%10027, USA}
%\affiliation{NEC-Laboratories America,
%Inc., 4~Independence Way, Princeton, NJ 085540 USA}

\chapter[D.~M.~Basko, I.~L.~Aleiner, and B.~L.~Altshuler]
{On the problem of many-body localization}%\label{BAA}
\begin{center}
 \textit{D.~M.~Basko, I.~L.~Aleiner, and B.~L.~Altshuler} \\
Physics Department, Columbia University, New York, NY 10027, USA
\end{center}

%\addcontentsline{toc}{chapter}%
%{\ref{BAA}~\textit{D.~M.~Basko, I.~L.~Aleiner, and B.~L.~Altshuler}%
%~~On the problem of many-body localization}

\begin{abstract}
We review recent progress in the study of transport properties of
interacting electrons subject to a disordered potential which is
strong enough to localize {\em all} single-particle states. This
review may also serve as a guide to the recent paper by the authors
[Annals of Physics (2006), in press]. Here we skip most of the
technical details and make an attempt to discuss the physical grounds
of the final-temperature metal-insulator transition described in the
above-mentioned paper.
\end{abstract}

\section{Introduction}

Transport properties of conducting materials at low temperature
$T$ are determined by an interplay between the interaction of the
itinerant electrons with each other and the quenched disorder
which creates a random potential acting on these electrons. In the
absence of the electron-electron interaction the most dramatic
phenomenon is Anderson
localization (Anderson, 1958) -- the dc electrical
conductivity $\sigma$ can be qualitatively different depending on
whether one-particle wave functions of the electrons are localized
or not. In the latter case $\sigma (T)$ has a finite
zero-temperature limit, while in the former case $\sigma (T)$
vanishes when $T \to 0$. Therefore, Anderson localization of
electronic states leads to the Metal to Insulator Transition at
zero temperature.

When discussing zero temperature conductivity~$\sigma(0)$, we
need to consider only electronic states close to the Fermi level.
The conductivity becomes finite at any finite temperature provided
that extended states exist somewhere near the Fermi level. It is
commonly accepted now that localized and extended states in a
random potential can not be mixed in the one-electron spectrum and
thus this spectrum in a general case is a combination of
bands of extended states and bands of localized states. A border
between a localized and an extended band is called mobility edge.
If the Fermi level is located inside a localized band and
inelastic scattering of the electrons is completely absent, the
conductivity should follow Arrhenius law
$\sigma(T)\propto\exp(-E_c/T)$, where $E_c$ is the distance from
the Fermi level to the closest mobility edge.  Another common
belief following from the scaling theory of Anderson localization
(Thouless, 1977; Abrahams {\em et al.}, 1979), is that in low
dimensionality $d$, namely at $d=1,2$ all states are localized in
an arbitrarily small disorder, while for free electrons (no
periodic potential) $E_c>0$ is finite at $d=3$.%
\footnote{
For $d=1$ this statement was proved rigorously both for one-channel
(Gertsenshtein and Vasil'ev, 1959; Berezinskii, 1973) and
multi-channel (Efetov and Larkin, 1983; Dorokhov, 1983) disordered
wires.}
It means that without inelastic processes 
$\sigma_{d=1,2}(T)=0$, while for $\sigma_{d=3}(T)$ one should
expect the Arrhenius law. Note that for electrons in a crystal
within a given conduction band the latter conclusion is not always
correct -- strong enough disorder can localize the whole band.

As soon as inelastic processes are included, the situation becomes
more complicated. In particular, electron-phonon interaction leads to
the mechanism of conductivity known as hopping conductivity (Fritzche,
1955; Mott, 1968a; Shklovskii and Efros, 1984) --  with an assistance
of phonons,
electrons hop between the localized states  without being activated
above the mobility edge. As a result, $\sigma(T)$ turns out to be
finite (although small) at arbitrarily low~$T$ even when all
one-electron states are localized.

Can interaction between electrons play the same role and cause the
hopping conductivity? This question was discussed in literature
for a long time (Fleishman and Anderson, 1980; Shahbazyan and Raikh,
1996; Kozub, Baranovskii, and Shlimak, 2000; Nattermann, Giamarchi,
and Le Doussal, 2003; Gornyi, Mirlin, and Polyakov, 2004) and no
definite conclusion was achieved. The problem is that although the
electric noise exists inside the material with a finite ac
conductivity\footnote{In this paper we mostly focus on dc
conductivity. As to ac conductivity, it never vanishes, because at
any frequency density of resonant pairs of states is finite.} the
``photons'' in contrast with phonons become localized together
with electrons.

In a recent work (Basko, Aleiner, and Altshuler, 2006) we have
demonstrated that
electron-electron interaction alone cannot cause finite
conductivity even when temperature is finite, but small enough. In
the absence of phonons and extended one-electron states conductivity
of a system of interacting electrons vanishes exactly below some
critical temperature $T_c$. At the same time, at high temperatures
$T>T_c$ the conductivity $\sigma (T)$ is finite. It means that at $T =
T_c$ the system of interacting electrons subject to a random potential
undergoes a genuine phase transition that manifests itself by the
emerging of a finite conductivity!

This transition can be thought of as many-body localization -- it
applies to many-body eigenstates of the whole system. This
localization occurs not in the real space, but rather in the Fock
space. This fact does not affect the validity of the concept of
mobility edge. In fact, the existence of the "metallic" state at
$T > T_c$ implies that the many-body states with energies
$\mathcal{E}$   above $\mathcal{E}_c$ are  extended. One can
estimate the difference between $\mathcal{E}_c$ and the energy of
the many-body ground state $\mathcal{E}_0$ as
$\mathcal{E}_c-\mathcal{E}_0 \sim T\mathcal{N}(T)$, where
$\mathcal{N}(T)$ is the total number of one-particle states in the
energy strip of the width $T$. Note that the existence of the
extended many-body states above the mobility edge does not
contradict the fact that below  $T_c$ there is no conductivity --
in contrast with the case of one-particle localization there is no
Arrhenius regime since $\mathcal{E}_c-\mathcal{E}_0$ turns out to
be proportional to the volume of the system,  {\em i.e.}, is
macroscopically large (see Sec.~\ref{sec:MBloc} below for more
details).

In order to avoid possible misunderstanding we would like to
emphasize that we focus only on the inelastic collisions between the
electrons, {\em i.e.}, on creation or annihilation of {\em real}
electron-hole pairs. There are other effects of electron-electron
interactions which can be understood as renormalization of the
one-particle random potential by the interaction. Being
temperature-dependent, this renormalization leads to a number of
interesting effects, such as the interaction corrections to the
density of states and conductivity in disordered metals (Altshuler and
Aronov, 1985). On
the insulating side of the one-particle localization transition
similar effects cause the well-known Coulomb gap (Shklovskii and
Efros, 1984)
which reduces hopping conductivity. On the other hand, this is just
a correction to the time-independent random potential. As such, it can
maybe shift the position of the many-body Metal to Insulator
transition, {\em i.e.}, renormalize $T_c$, but is unable to
destabilize the insulating or metallic phases. From now on we will
simply neglect all elastic (Hartree-Fock) effects and concentrate on
the {\em real} inelastic electron-electron collisions.

Localization of the many-body states in the Fock space has been
discussed by Altshuler {\em et al.} (1997) for the case
of zero-dimensional systems with finite, although large, number of
electrons. In this paper the authors proposed an approximate mapping
of the Hamiltonian of a metallic grain with large Thouless conductance
$g$ and moderate interaction between the electrons to the
one-particle Hamiltonian on a lattice with the topology of the
Cayley tree and an on-site disorder. The latter problem has an
exact solution (Abou-Chacra, Anderson, and Thouless, 1973; Efetov,
1987) that exhibits the localization
transition. In terms of interaction electrons this transition
means that one-particle excitation states below certain energy are
quite close to some exact many-body 
excitations. As to the one-particle excitations with energies
higher than the critical one, its wave function can be viewed as a
linear combination of a large number of the many-body
eigenstates.

For an infinite system ($d>0$) the situation is more complex, and
Cayley tree approximation is hard to justify. Nevertheless,
a consistent analysis of a model with weak and short range
interaction to all orders of perturbation theory enabled us to
analyze the many-body localization transition and to demonstrate
that both the metallic state at high temperatures and the
insulating state at low temperatures are stable and survive all
higher loop corrections to the locator expansion. Therefore, the
existence of the transition is proved on the physical level of rigor.

It should be noted that such an insulating state that is
characterized by exactly zero conductivity is quite different from
all other known types of insulators. For example, Mott insulator is
believed to have finite, though exponentially small conductivity at
finite temperatures.

The present text represents a shortened version of the paper by Basko,
Aleiner, and Altshuler (2006), hereafter referred to as BAA~paper. We
omit most of the technical details (for which the reader will be
referred to specific sections of the BAA~paper), and stress the key
ideas.

The remainder of the paper is organized as follows. In
Sec.~\ref{sec:background} we briefly review some well-known facts
about electric conduction in Anderson insulators and pose the
problem. Sec.~\ref{sec:solution} represents a sketch of the
solution whose details are given in BAA~paper. We discuss the
model for interacting localized electrons in Sec.~\ref{sec:model},
and the corresponding Fock space picture in Sec.~\ref{sec:Fock}.
In Sec.~\ref{sec:statistics} we show the formal way to
characterize metallic and insulating phases. In
Sec.~\ref{sec:SCBA} we introduce the main approximation used in
the calculation (self-consistent Born approximation), and discuss
its validity. The existence of the metallic state at high
temperatures and its properties are discussed in
Sec.~\ref{sec:metal}. Sec.~\ref{sec:insulator} is dedicated to the
proof of existence of the insulating phase at low
temperatures; the value of the transition temperature is obtained
as the limit of stability of the insulating phase. In
Sec.~\ref{sec:MBloc} we discuss the macroscopic implications of
the problem, introducing the concepts of many-body localization
and many-body mobility edge. Finally, in
Sec.~\ref{sec:conclusions} we summarize the results and present an
outlook of the future developments.

\section{Background and formulation of the problem}
\label{sec:background}

\subsection{Non-interacting electrons in disorder potential}

Let us briefly review the basic concepts developed for the problem of
one-electron wave functions in a disordered potential in $d$
dimensions. Depending on the strength of the disorder potential, a
wave function $\phi_\alpha(\vec{r})$ of an eigenstate~$\alpha$ with
the energy~$\xi_\alpha$ can be either {\em localized} or {\em
  extended}:
\begin{equation}
|\phi_\alpha(\vec{r})|^2\propto
\left\{
\begin{array}{ll}
\frac{1}{\zeta_{loc}^d}
\exp\left(-\frac{|\vec{r}-\vec\rho_\alpha|}{\zeta_{loc}}\right),
& {\rm localized;}\\
\frac{1}{\Omega}, & {\rm extended.}
\end{array}\right.
\label{eq:2.1}
\end{equation}
Here $\zeta_{loc}$ is the localization length which depends on the
eigenenergy $\xi_\alpha$, and $\Omega$ is the volume of the
system. Each localized state is characterized by a point in space,
$\vec\rho_\alpha$, where $|\phi_\alpha(\vec{r})|^2$ reaches its
maximum, and an exponentially falling envelope. Extended states
spread more or less uniformly over the whole volume of the
system. Localized and extended states cannot coexist at the same
energy, and the spectrum splits into bands of localized and extended
states. The energies separating such bands are known as {\em mobility
  edges}. For free electrons in $d\geq{3}$ disorder potential leads to
only one mobility edge $\mathcal{E}_1$, so that
\begin{eqnarray}
\xi_\alpha < \mathcal{E}_1: &\quad {\rm localized};
\nonumber
\\
\xi_\alpha > \mathcal{E}_1: &\quad  {\rm extended}.
\label{eq:2.2}
\end{eqnarray}
If a finite mobility edge~(\ref{eq:2.2}) exists and the Fermi
level~$\epsilon_F$ lies in the band of localized states, the
conductivity is determined by the exponentially small occupation
number of the delocalized states 
\begin{equation}
\sigma(T)\propto{e}^{-(\mathcal{E}_1-\epsilon_F)/T}.
\label{Arrhenius}
\end{equation}

In this paper we are interested in transport properties of the
systems where {\em all} single-particle states are localized, and thus
without many-body effects $\sigma=0$ at any temperature. It is well
established now that the mobility edge usually does not exist for
one- and two-dimensional systems, and all single-particle states are
indeed localized for an arbitrarily weak disorder. Such a situation
can arise for a large~$d$ as well, if the bandwidth is finite
and disorder is sufficiently strong.

\subsection{Role of inelastic processes and phonon-assisted hopping}
\label{sec:inelastic}

As long as all single-particle states are localized, transport occurs
only because of inelastic processes, which transfer electrons between
different localized eigenstates. At this stage we introduce the main
energy scale of the problem: the typical energy spacing between states
whose spatial separation does not exceed~$\zeta_{loc}$, so that there
is overlap between their wave functions:
\begin{equation}
\delta_\zeta=\frac{1}{\nu\zeta_{loc}^d}\,,
\end{equation}
where $\nu$~is the one-particle density of states per unit volume.

The conductivity is, roughly speaking, proportional to the rate of the
transitions between different localized states, which can be called
inelastic relaxation rate~$\Gamma$. Obviously, at $T=0$ inelastic
processes disappear, so regardless of the mechanism of inelastic
relaxation the conductivity must vanish:
\begin{equation}
\lim_{T\to{0}}\sigma(T)=0\,.
\end{equation}
The question is how $\sigma(T)$ approaches zero for each particular
mechanism.

When phonons are the main source of inelastic scattering, the answer
is given by Mott's variable range hopping
formula (Mott, 1968a)\footnote{Here we do not
  consider effects of the Coulomb interaction which are known to
  modify the power of temperature in the exponent (Shklovskii and
  Efros, 1984).}
\begin{equation}\label{MottVRH=}
\sigma(T)=\sigma_0\left(\frac{T}{\delta_\zeta}\right)^\alpha
\exp\left[-\left(\frac{\delta_\zeta}{T}\right)^{1/(d+1)}\right],
\end{equation}
where $\sigma_0$~and~$\alpha$ are constants. According to
Eq.~(\ref{MottVRH=}), $\sigma(T)$ remains finite as long as
$T\neq{0}$. The reason for $\sigma(T)\neq{0}$ is that for any pair of
localized states one can always find a phonon whose frequency exactly
corresponds to their energy mismatch, at low temperature one should
just wait long enough.

The same type of $\sigma(T)$-dependence would result from the coupling
of electrons with {\em any} delocalized thermal bath whose energy
spectrum is continuous down to zero energy. The specific nature of the
bath at most affects the power-law prefactor. On the contrary, the
stretched exponential factor is universal; it originates from the
counting of electronic states available for the transition, and does
not depend on the specific scattering mechanism.

\subsection{Inelastic relaxation due to electron-electron interaction}

Now let us assume that there is no external bath coupled to electrons, 
but some electron-electron interaction is present. What will be the
dependence of~$\sigma(T)$? In line with the discussion of
the previous subsection, the question should be posed as follows: do
electron-hole pairs themselves provide a suitable bath in a localized
system, thus validating Eq.~(\ref{MottVRH=})?

One possible answer is ``yes''. Indeed, recall Mott formula for the
low-tempera\-ture dissipative ac conductivity $\sigma(\omega)$ in a
localized system (Mott, 1968b):
\begin{equation}\label{MottAC=}
\sigma(\omega)=\sigma_1\,\frac{\omega^2}{\delta_\zeta^2}
\ln^{d+1}\frac{\delta_\zeta}{|\omega|}\,.
\end{equation}
According to fluctuation-dissipation theorem, at finite temperature
electromagnetic fluctuations of a finite spectral density should be
present, and they might serve as a bath. The problem with this
argument is that Eq.~(\ref{MottAC=}) is the {\em spatial average} of
the conductivity over the whole (infinite) volume. For each given
realization of disorder, excitations determining $\sigma(\omega)$ from
Eq.~(\ref{MottAC=}) are localized, and although the total volume is
infinite, the spectrum of electron-hole pairs is effectively discrete.

The crucial point is that as long as electron-electron interaction is
local in space (here we do not consider long-range interactions), it
effectively couples electronic states only within the same
localization volume, where the spectrum of electronic states is
effectively discrete. The following sections are dedicated to a
systematic discussion of this problem which was first pointed out by
Fleishman and Anderson (1980). The conclusion can be stated as
follows: electron-hole excitations can cause finite conductivity only
if the temperature of the system exceeds some critical value. At lower
temperatures $\sigma(T)$ vanishes exactly.

\section{Finite-temperature metal-insulator transition}
\label{sec:solution}

\subsection{Matrix elements of electron-electron interaction between localized
states: essential features of the model} \label{sec:model}

For simplicity we consider a system of spinless electrons and assume
that electron-electron interaction is weak and short-range:%
\footnote{Interaction proportional to $\delta(\vec{r}_1-\vec{r}_2)$ in
  the strict sense is equivalent to no interaction for spinless
  electrons, considered here, due to the Pauli principle. Here, by
  writing $\delta(\vec{r}_1-\vec{r}_2)$ we only mean that the range is
  much smaller than the electron mean free path, so it is not a true
  $\delta$-function.}
\begin{equation}\label{Vr1r2=}
V(\vec{r}_1-\vec{r}_2) =\displaystyle\frac\lambda\nu\,
\delta(\vec{r}_1-\vec{r}_2)\,,
%\quad\frac{1}\nu\equiv \delta_\zeta\,\zeta_{loc}^d\,,
\end{equation}
where $\lambda\ll{1}$ is the dimensionless interaction constant,
$\nu$~is the one-particle density of states per unit volume.

In the basis of localized single-particle eigenstates the hamiltonian
corresponding to the pair interaction potential~(\ref{Vr1r2=}) takes
the form
\begin{equation}\label{Hloc=}
\hat{H}=\sum_{\alpha}\xi_\alpha
\hat{c}_\alpha^\dagger\hat{c}_\alpha
+\frac{1}2\sum_{\alpha\beta\gamma\delta} V_{\alpha\beta\gamma\delta}\,
\hat{c}_\alpha^\dagger\hat{c}_\beta^\dagger
\hat{c}_\gamma\hat{c}_\delta
\end{equation}
Consider the structure of the matrix elements
$V_{\alpha\beta\gamma\delta}$. Since $V(\vec{r})$ is short-range, they
decrease exponentially when the spatial separation between the states
increases, the characteristic scale being the localization length
$\zeta_{loc}$. In addition to this spatial suppression, the matrix
elements decrease rapidly when the energy difference, say
$\xi_\alpha-\xi_\gamma$, increases exceeding the level spacing
$\delta_\zeta$. This occurs because the localized wave functions
oscillate randomly, and the bigger the energy difference, the weaker
are these random oscillations correlated (Altshuler and Aronov, 1985).
Provided that the restrictions
\begin{eqnarray}
&|\vec{r}_\alpha-\vec{r}_\beta|\lesssim\zeta_{loc},\quad
|\vec{r}_\alpha-\vec{r}_\gamma|\lesssim\zeta_{loc},\quad
|\vec{r}_\beta-\vec{r}_\gamma|\lesssim\zeta_{loc},\quad
\mathrm{etc.},&\label{spatialcutoff=}\\
&|\xi_\alpha-\xi_\delta|,|\xi_\beta-\xi_\gamma|\lesssim\delta_\zeta
\quad{\rm or}\quad
|\xi_\alpha-\xi_\gamma|,|\xi_\beta-\xi_\delta|\lesssim
\delta_\zeta,& \label{energycutoff=}
\end{eqnarray}
are fulfilled, we have
$|V_{\alpha\beta\gamma\delta}|\sim\lambda\delta_\zeta$.

There are several ways to model these essential properties of
$V_{\alpha\beta\gamma\delta}$. In BAA~paper a specific model was
adopted;
essentially, the space and energy dependences of the matrix elements
were replaced by simple rectangular cutoffs (see Sec.~3 of
BAA~paper for details).

\subsection{Many-electron transitions and Fock space}
\label{sec:Fock}

We wish to note that all the discussion of this subsection is not
conceptually new. In fact, it is just a generalization of the
arguments of Altshuler {\em et al.} (1997) to an infinite system.

Conventionally, an elementary inelastic process is a decay of one
single-particle excitation (an electron occupying a state~$\alpha$)
into three single-particle excitations -- a hole in the state~$\beta$
and two electrons in the states $\gamma$~and~$\delta$. Such a decay
can be described differently: one can say that the hamiltonian couples
the single-particle excitation with the three-particle excitation by
the matrix element $V_{\alpha\beta\gamma\delta}$. Further action of
the interaction hamiltonian produces five-particle excitations,
seven-particle excitations, {\em etc.}:
\begin{equation}
\xi_\alpha\;\rightarrow\;
\xi_\gamma+\xi_\delta-\xi_\beta\;\rightarrow\;
\xi_1+\xi_2+\xi_3-\xi_4-\xi_5\; \rightarrow\ldots\,.
\end{equation}
If on each stage the coupling is strong enough ({\em i.~e.}, the
matrix element is of the same order or larger than the corresponding
energy mismatch), the single-particle excitation indeed decays
irreversibly into all possible many-body states. In other words, exact
many-body eigenstates become delocalized in the Fock space. If,
oppositely, three-particle states contribute only a weak perturbative
admixture to the one-particle state, the contribution from
five-particle states is even weaker, {\em etc.}, the initial electron
will never decay completely. One can say that it is localized in the
Fock space.

Another way to visualize the inelastic relaxation is to look at the
energy structure of the quasiparticle spectral function:
\begin{equation}
A_\alpha(\epsilon)=\sum_{k} \left|\langle\Psi_k|
\hat{c}^\dagger_\alpha|\Psi_0\rangle\right|^2 \delta(\epsilon+E_0-E_k).
\end{equation}
Here $\Psi_0$ and $\Psi_k$ are many-body eigenstates ($\Psi_0$ is not
necessarily a ground state),
$E_0$~and~$E_k$ are the corresponding energies. Basically,
$A_\alpha(\epsilon)$ shows how the single-particle excitation on top
of a given eigenstate is spread over other many-body eigenstates of
the system.

\begin{figure}
\includegraphics[width=\textwidth]{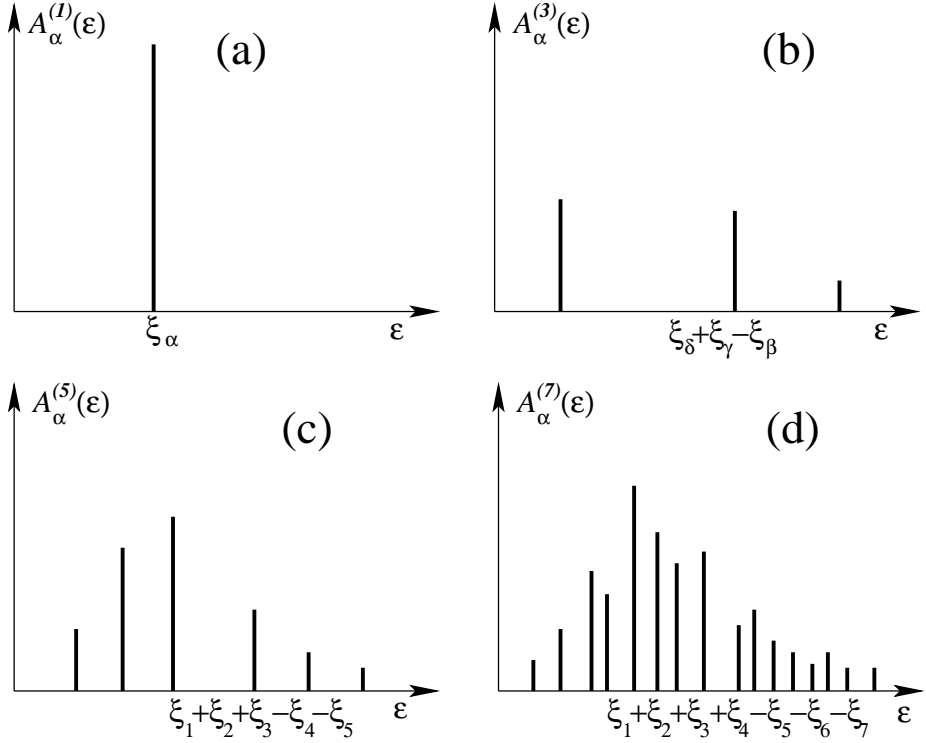}
\caption{A schematic view of the
contributions to the perturbative
expansion~(\ref{spectralperturbative=}) of the spectral function
$A_\alpha(\epsilon)$ from one-particle, three-particle,
five-particle, and seven-particle
excitations.}\label{fig:spectral}
\end{figure}

One can represent the spectral function in the form of expansion in
powers of the interaction constant~$\lambda$
(Fig.~\ref{fig:spectral}):
\begin{equation}
A_\alpha(\epsilon)=\sum_{n=0}^\infty\lambda^{2n}A_\alpha^{(2n+1)}(\epsilon).
\label{spectralperturbative=}
\end{equation}
$A_\alpha^{(1)}(\epsilon)$ corresponds to the bare quasiparticle peak,
$\delta(\epsilon-\xi_\alpha)$ . Linear in~$\lambda$ term represents a
random Hartree-Fock shift of the energy~$\xi_\alpha$ which is already
assumed to be random. This linear term is of no interest to us,
so it is not included in the
expansion~(\ref{spectralperturbative=}). The $\lambda^2$ term
corresponds to the contribution of three-particle excitations. The
number of these excitations is effectively finite due to the
restrictions~(\ref{spatialcutoff=}),~(\ref{energycutoff=}), so
$A^{(3)}(\epsilon)$ is a collection of $\delta$-peaks at
three-particle energies $\xi_\gamma+\xi_\delta-\xi_\beta$. Again,
$\lambda^3$~terms correspond only to peak shifts and are omitted. The
five-particle contribution $A^{(5)}(\epsilon)$ is again a collection
of $\delta$-peaks, however, spaced more closely than three-particle
peaks. This represents the general rule: the more particles are
involved, the higher is the density of the peaks. At the same time,
contributions of many-particle processes are suppressed due to the
smallness of~$\lambda$. What is crucial to us, is the result of this
competition at $n\rightarrow\infty$: will the empty spaces between
peaks be filled making $A_\alpha(\epsilon)$ a continuous function, or
the process will be suppressed by the powers of~$\lambda$ and
$A_\alpha(\epsilon)$ will remain a collection of $\delta$-peaks?

Thus, the problem of inelastic relaxation of a quasiparticle due to
electron-electron interaction is related to the problem of localization
or delocalization of excitations in the many-electron Fock space.
The simplest model describing localization-delocalization physics is
Anderson model. Anderson (1958) considered a tight-binding
model on a $d$-dimensional lattice. The coupling between sites is
nearest-neighbor only with a fixed matrix element~$V$. The on-site
energies are assumed to be random and uncorrelated, with some typical
value denoted by~$W$.

It is thus tempting to identify the many-electron
hamiltonian~(\ref{Hloc=}) with the Anderson hamiltonian on a certain
lattice, whose sites correspond to many-particle
excitations~\cite{AGKL}. The approximate rules of correspondence
then should be the following:
\begin{itemize}
\item
$V\rightarrow\lambda\delta_\zeta$ -- typical value of the coupling
  matrix element;
\item
$W\rightarrow\delta_\zeta$ -- typical energy mismatch in each
  consecutive virtual transition,
  $|\xi_\alpha+\xi_\beta-\xi_\gamma-\xi_\delta|\sim\delta_\zeta$;
\item
finally, the coordination number $2d\rightarrow{T}/\delta_\zeta$. This
represents the number of three-particle excitations to which a
given single-particle excitation is coupled, with the energy mismatch
not exceeding~$\delta_\zeta$. The value $T/\delta_\zeta$ is obtained
as the product of the number of electrons~$\beta$ within the same
localization volume, available for the collision with the probe
electron~$\alpha$ ($\sim{T}/\delta_\zeta$), and the number of ways to
distribute energy allowed by the restriction~(\ref{energycutoff=}),
which is $\sim{1}$.
\end{itemize}
According to Anderson (1958),
localization-delocalization transition occurs at
\begin{equation}\label{Anderson=}
\frac{Vd}{W}\ln\frac{W}{V}\sim
\frac{\lambda{T}}{\delta_\zeta}\ln\frac{1}\lambda\sim{1}\,.
\end{equation}
For the interacting localized electrons this would imply that below
some temperature~$T_c$ the inelastic relaxation is frozen, so that the
conductivity vanishes exactly; above~$T_c$ some inelastic relaxation
is taking place , and the conductivity is finite. This corresponds to
a finite-temperature metal-insulator transition.

This analogy, however, should be used with caution. Anderson's result,
in fact, can be sensitive to the structure of the lattice; possible
sources of problems are listed in Sec.~2.2 of BAA~paper. One can
develop a systematic approach to the problem, based on the
diagrammatic technique for interacting electrons (Basko, Aleiner, and
Altshuler, 2006). Below we
headline the basic ideas of this approach to justify the assumption
that Eq.~(\ref{Anderson=}) indeed correctly determines the point of
the metal-insulator transition.

\subsection{Statistics of the transition rates}
\label{sec:statistics}

We have to admit, that all the discussion of this subsection is not
conceptually new either. It rather represents a generalization of the
arguments of Anderson (1958) to a many-body problem.

As discussed in previous sections, the focus of the problem is the
inelastic quasiparticle relaxation, which is represented by the
imaginary part of the single-particle self-energy:
\begin{equation}
\Gamma_\alpha(\epsilon)=\mathop{\mathrm{Im}}\Sigma^A_\alpha(\epsilon)\,.
\end{equation}
We stress that $\Gamma_\alpha(\epsilon)$ is a random quantity, as it
depends on the positions of all single-particle levels $\{\xi_\beta\}$
and all occupation numbers $\{n_\beta\}$. One has no other choice but
to perform statistical analysis of this random quantity.

How to distinguish between metallic and insulating phases within a
statistical framework? It is clear that positions of the peaks in
$\Gamma(\epsilon)$ for the insulating regime wander randomly with the
variation of random energies $\xi_\alpha$ from Eq.~(\ref{Hloc=}). Due
to this variation, the ensemble average of the decay rate
$\langle\Gamma(\epsilon)\rangle$ is the same in both phases and can
not be used for the distinction. In fact, one has to investigate the
whole distribution function $P(\Gamma)$.

\begin{figure}
\includegraphics[width=\textwidth]{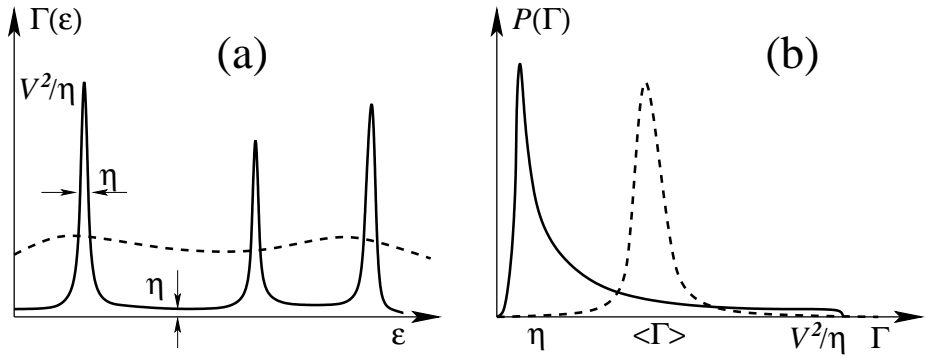}
\caption{ (a)~Schematic energy
dependence of the quasiparticle decay rate $\Gamma(\epsilon)$ in
the metallic (dashed line) and insulating (solid line) phases for
a given realization of single-particle levels $\{\xi_\alpha\}$.
(b)~The corresponding distribution functions $P(\Gamma)$ for
$\Gamma(\epsilon)$ at a given energy~$\epsilon$ in the metallic
(dashed line) and insulating (solid line) phases.} \label{figP}
\end{figure}

To understand the difference in the behavior of $P(\Gamma)$ in the
two phases, it is instructive to start with the behavior of
$\Gamma_\alpha(\epsilon)$ for a given realization of disorder,
Fig.~\ref{figP}a. Deep in the metallic phase
$\Gamma_\alpha(\epsilon)$ is a smooth function of energy, so its
distribution $P(\Gamma)$ at a given energy is a narrow gaussian.
In the insulating phase $\Gamma_\alpha(\epsilon)$ is given by a
sequence of infinitely narrow $\delta$-peaks, so an arbitrarily
chosen value of~$\epsilon$ falls between the peaks with the
probability~1, giving $\Gamma=0$. If $\epsilon$ hits a
$\delta$-peak, then $\Gamma=\infty$, which happens, however, with
zero probability.

One can deal with this uncertainty by introducing an
infinitesimal imaginary energy shift (damping)~$\eta$,
%$\epsilon\rightarrow\epsilon+i\eta$, 
and consider $\Gamma_\alpha(\epsilon+i\eta)$. Physically, it may be
viewed as an infinitesimally weak coupling to a dissipative bath
(e.~g., phonons). Clearly, in the metallic phase it has no effect,
while in the insulating phase this damping broadens $\delta$-peaks
into lorentzians of the width~$\eta$, which leads to appearance of the
tails. Now, even if the energy~$\epsilon$ falls between the peaks,
$\Gamma$~will have a finite value proportional to~$\eta$. As a result,
the distribution function $P(\Gamma)$ will have the form sketched in
Fig.~\ref{figP}b. Now, having calculated the distribution function for
a small but finite~$\eta$, one can distinguish between the metallic
and the insulating phases according to
\begin{equation}
\lim_{\eta\to 0}\lim_{\Omega\to \infty}
P(\Gamma>0)\;
\left\{\begin{array}{ll}
>0,\;\;\; & {\rm metal}\\
=0, & {\rm insulator.}
\end{array}
\right. \label{criterion}
\end{equation}
Here $\Omega$ is the total volume of the system. We emphasize
that the order of limits in Eq.~(\ref{criterion}) cannot be
interchanged, since in a finite closed system the spectrum always
consists of discrete $\delta$-peaks.

\subsection{Self-consistent Born approximation}
\label{sec:SCBA}

Our main object of interest
is the probability distribution function of the decay rate
$P(\Gamma)$. Its calculation is performed in two stages: (i)~we find
$\Gamma_\alpha(\epsilon)$ for a given realization of disorder,
(ii)~we calculate its statistics. This subsection is dedicated to the
first task.

We intend to describe both metallic and insulating regimes. In the
latter regime relaxation dynamics is absent, the system never reaches
thermal equilibrium, and temperature itself is ill-defined. Therefore,
the only appropriate formal framework is the non-equilibrium
formalism of Keldysh (1964) (see Sec.~4.1 of BAA~paper for details).

\begin{figure}
\includegraphics[width=0.7\textwidth]{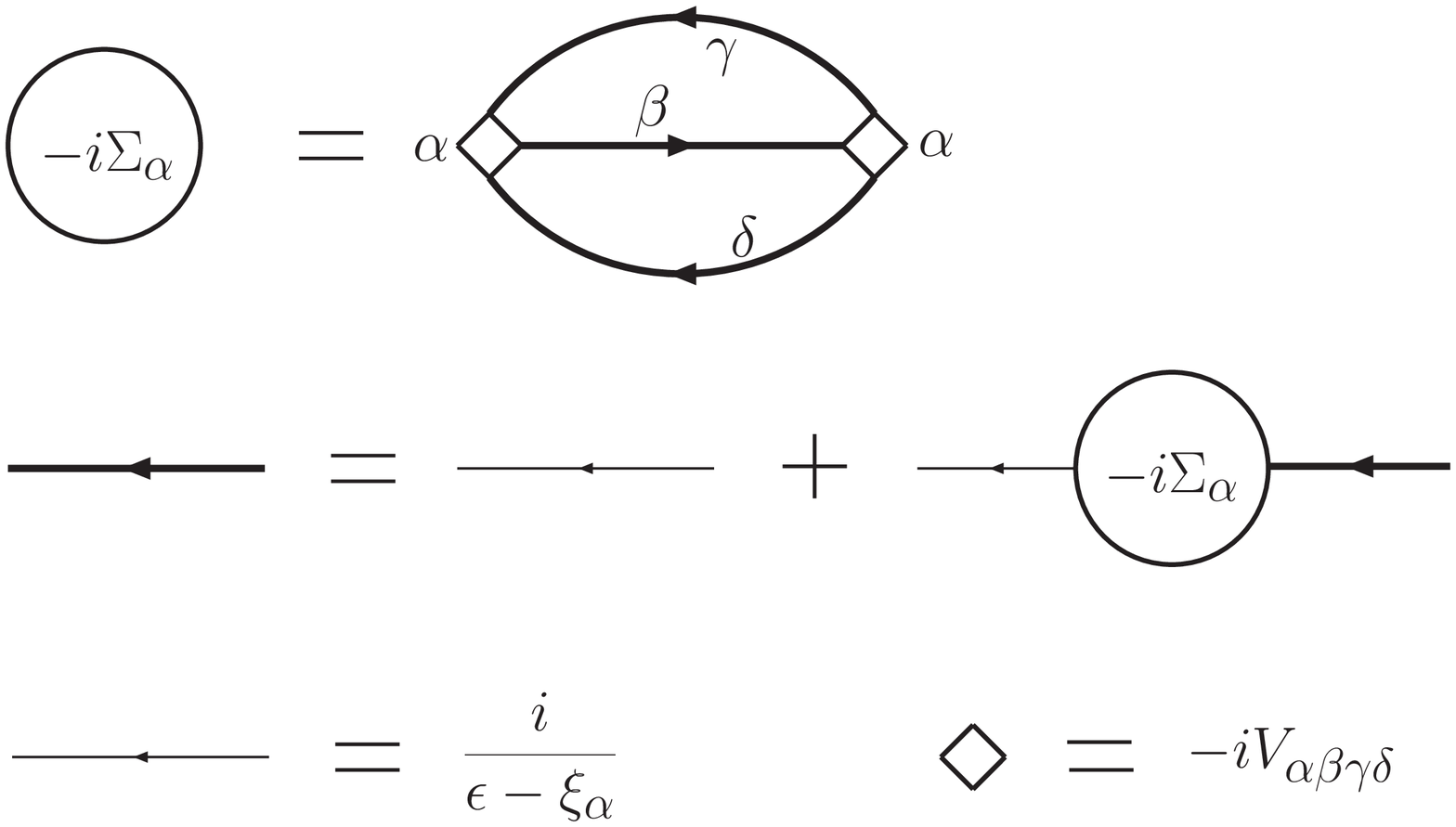}
\caption{Diagrammatic representation of self-consistent Born
approximation.}\label{fig:SCBA}
\end{figure}

Our approach for calculation of $\Gamma_\alpha(\epsilon)$ is the
self-consistent Born approximation (SCBA), shown diagrammatically in
Fig.~\ref{fig:SCBA}, and given by
\begin{eqnarray}
\Gamma_\alpha(\epsilon)&=&\eta+\pi\sum_{\beta,\gamma,\delta}
|V_{\alpha\beta\gamma\delta}|^2\int d\epsilon'\,d\omega
\,A_\beta(\epsilon')\, A_\gamma(\epsilon'+\omega)\,
A_\delta(\epsilon-\omega)\times\nonumber\\ &&\qquad\qquad\times
\left[n_\beta(1-n_\gamma)(1-n_\delta)+(1-n_\beta)n_\gamma{n}_\delta\right],
\label{SCBArate=}\\
A_\alpha(\epsilon)&=&\frac{1}{\pi}\,\frac{\Gamma_\alpha(\epsilon)}
{(\epsilon-\xi_\alpha)^2+\Gamma_\alpha^2(\epsilon)}\,.
\label{SCBAspectral=}
\end{eqnarray}
Here $n_\alpha=0,1$ is the fermion occupation number of the
single-particle state~$\alpha$. It is crucial to realize that one
cannot replace~$n_\alpha$ with its average equilibrium value,
corresponding to Fermi-Dirac distribution. The basis in the Fock space
is formed by Slater determinants corresponding to $n_\alpha=0,1$
only, and $\Gamma$~represents the rate of a transition between two
such basis states with different sets of $\{n_\alpha\}$. Temperature
enters through total energy of the state, which is determined by
all~$\{n_\alpha\}$ and is proportional to the volume of the system
(this issue will be discussed in more detail in Sec.~\ref{sec:MBloc}
of the present paper).

Iterations of SCBA generate self-energy diagrams which have one common
property: they describe decay processes where the number of particles
in the final state is maximized for each given order
in~$\lambda$, thus maximizing the phase space available for the
decay. Vice versa, each diagram satisfying this condition is taken
into account by SCBA. Let us briefly discuss contributions which are
neglected by Eqs.~(\ref{SCBArate=}),~(\ref{SCBAspectral=}); for a
detailed discussion see Sec.~7 of BAA~paper.
\begin{itemize}
\item
Eqs.~(\ref{SCBArate=}),~(\ref{SCBAspectral=}) completely ignore the
real part of the self-energy, $\mathop{\mathrm{Re}}\Sigma$. In most
of the terms effect of $\mathop{\mathrm{Re}}\Sigma$ reduces to random
uncorrelated corrections to already random energies, and appear to be
completely negligible. In a more accurate approximation, however, some
statistical correlations are present. They renormalize the numerical
prefactor in the expression for the critical temperature, see Sec.~7.3
of BAA~paper.
\item
Generally, quantum mechanical probability of a transition is given by
a square of the total amplitude, the latter being a sum of partial
amplitudes. In fact, Eqs.~(\ref{SCBArate=}),~(\ref{SCBAspectral=})
correspond to replacing the square of the sum by the sum of the
squares. Approximation which neglects the interference terms can be
justified in the same way as in the Anderson model of a high
dimensionality $d_{eff}\sim{T}/\delta_\zeta\sim{1}/\lambda\gg{1}$, see
Sec.~7.2 of BAA~paper.
\item
Finally, diagrams generated by SCBA correspond to taking all
$n$-particle vertex function in the leading approximation in~$\lambda$.
One can show that for $\lambda\ll{1}$ vertex corrections are small,
see Sec.~7.1 of BAA~paper. For $\lambda\sim{1}$ one would have to
introduce the full $n$-particle vertex analogously to how it is done
in conventional Fermi liquid  theory for $n=2$ (Landau, 1958;
Eliashberg, 1962). In this case the
result for the transition temperature will be determined not by the
bare interaction constant, but by the statistics of the full vertex
functions. Existence and regularity of these vertex functions are
basically equivalent to the assumption that in the absence of disorder
the interacting system is a Fermi liquid.
\end{itemize}

\subsection{Metallic phase}
\label{sec:metal}

Starting from the self-consistent equations
(\ref{SCBArate=}),~(\ref{SCBAspectral=}), one can straightforwardly
verify (see Sec.~5.1 of BAA~paper) that $P(\Gamma)$ can be well
approximated by gaussian distribution with the average and dispersion
\begin{equation}
\langle\Gamma\rangle\sim\lambda^2T,\quad
\langle\Gamma^2\rangle-\langle\Gamma\rangle^2\sim
\lambda^2\delta_\zeta^2,
\end{equation}
provided that
\begin{equation}\label{developedmetal=}
\sqrt{\langle\Gamma^2\rangle-\langle\Gamma\rangle^2}
\ll\langle\Gamma\rangle
\;\;\;\Leftrightarrow\;\;\;
T\gg{T}^{(in)}\sim\frac{\delta_\zeta}\lambda.
\end{equation}
According to the arguments of Sec.~\ref{sec:statistics}, this is
characteristic of the metallic phase. One should not think, however,
that this automatically means that the system has the same transport
properties  as conventional metals (which would mean that the
conductivity is given by the Drude formula). The system conducts in
the Drude regime only when the inelastic processes completely destroy
the localization, the latter manifesting itself as the weak
localization correction to conductivity (Altshuler and Aronov,
1985). This occurs when the discrete levels are completely smeared:
\begin{equation}
\langle\Gamma\rangle\gg\delta_\zeta\;\;\;\Leftrightarrow\;\;\;
T\gg{T}^{(el)}\sim\frac{\delta_\zeta}{\lambda^2}.
\end{equation}
It turns out that for $\lambda\ll{1}$ there is a parametric range of
temperatures:
\begin{equation}
\frac{\delta_\zeta}{\lambda}\ll{T}\ll\frac{\delta_\zeta}{\lambda^2}.
\end{equation}
In this interval electron-electron interaction is already sufficient
to cause inelastic relaxation, however, the localized nature of the
single-particle wave functions remains important. Conduction in this
regime can be viewed as an analog of hopping conduction discussed in
Sec.~\ref{sec:inelastic}, in the sense that electrons themselves
indeed provide a good bath. However, as $T\gg\delta_\zeta$, there is
no exponential factor in the temperature dependence $\sigma(T)$. This
dependence can be obtained from the kinetic equation describing
electron transitions between localized states, and is given by a power
law, $\sigma(T)\propto{T}^\alpha$, where $\alpha$~can be
model-dependent. Also, in this temperature range Wiedemann-Frantz law
can be violated. For further details the reader is referred to
Sec.~5.2 of BAA~paper. We only note that this regime is somewhat
analogous to the phonon-assisted conduction discussed by Gogolin,
Mel'nikov, and Rashba (1975).

\subsection{Insulating phase}
\label{sec:insulator}

Self-consistent equations
(\ref{SCBArate=}),~(\ref{SCBAspectral=}) represent a system of
nonlinear integral equations whose coefficients are random due to
randomness of level energies $\{\xi_\alpha\}$ and occupation
numbers $\{n_\alpha\}$. Apparently, for $\eta=0$ these equations
have a solution $\Gamma_\alpha(\epsilon)=0$, corresponding to the
insulating phase. One must check, however, whether this
solution is stable with respect to an infinitesimal
damping~$\eta$.

To perform the standard linear stability analysis of
Eqs.~(\ref{SCBArate=}),~(\ref{SCBAspectral=}) we linearize
Eq.~(\ref{SCBAspectral=}) as
\begin{equation}
A_\alpha(\epsilon)=\delta(\epsilon-\xi_\alpha)+
\frac{1}{\pi}\,\frac{\Gamma_\alpha(\epsilon)}
{(\epsilon-\xi_\alpha)^2}+O(\Gamma^2)\,,
\end{equation}
in complete analogy with Abou-Chacra, Anderson, and Thouless (1973),
substitute this
into Eq.~(\ref{SCBArate=}), and obtain a linear integral equation:
\begin{eqnarray}\nonumber
\Gamma_\alpha(\epsilon)=\eta+\sum_{\beta,\gamma,\delta}
|V_{\alpha\beta\gamma\delta}|^2\,
\frac{2\Gamma_\gamma(\epsilon+\xi_\beta-\xi_\delta)
+\Gamma_\beta(\xi_\gamma+\xi_\delta-\epsilon)}
{(\epsilon+\xi_\beta-\xi_\gamma-\xi_\delta)^2}\times\\ \times
\left[n_\beta(1-n_\gamma)(1-n_\delta)+(1-n_\beta)n_\gamma{n}_\delta\right].
\label{linSCBA=}
\end{eqnarray}
The solution of this equation may be sought in the form of a
perturbations series
\begin{equation}
\Gamma_\alpha(\epsilon)=\sum_{n=0}^\infty
\Gamma_\alpha^{(n)}(\epsilon)\,,
\end{equation}
where $\Gamma^{(n)}$ is of the order
$|V|^{2n}\sim(\lambda\delta_\zeta)^{2n}$ and is obtained after
$n$~iterations of Eq.~(\ref{linSCBA=}) starting from
$\Gamma^{(0)}=\eta$. Each term in this expansion can be calculated
explicitly and its statistics can be determined (details of this
rather cumbersome calculation are given in Sec.~6 of
BAA~paper). The resulting probability distribution function
is controlled by a single parameter~$\gamma_n$, which determines
the typical scale of the random quantity~$\Gamma^{(n)}$:
\begin{equation}
P_n(\Gamma^{(n)})=e^{-\gamma_n/\Gamma^{(n)}}
\sqrt{\frac{\gamma_n/\pi}{\left[\Gamma^{(n)}\right]^3}}\,,\quad
\gamma_n=C_1\eta\Lambda^{2n}\,,\quad
\Lambda\equiv
C_2\,\frac{\lambda{T}}{\delta_\zeta}\ln\frac{1}\lambda.
\label{result=}
\end{equation}
Here $C_1\sim{1}$ and $C_2\sim{1}$ are model-depndent numerical
constants (Eq.~(172) of BAA~paper
contains their values for the specific model adopted there).

\begin{figure}
\includegraphics[width=0.8\textwidth]{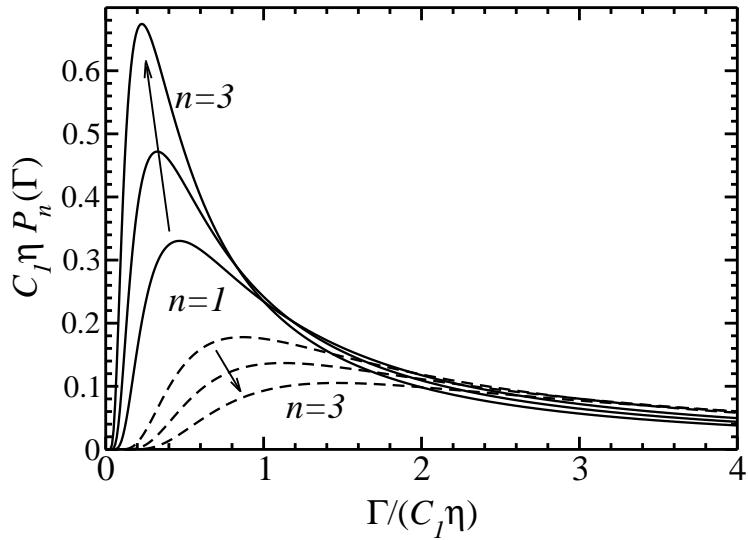}
\caption{Plot of the distribution functions $P_n(\Gamma^{(n)})$ as
given by Eq.~(\ref{result=}) for $n=1,2,3$ for $T/T_c=0.7$ (solid
lines) and $T/T_c=1.3$ (dashed lines).}\label{fig:distribution}
\end{figure}

The behavior of this distribution function at $n\rightarrow\infty$
is qualitatively different, depending on whether $\Lambda$~in
Eq.~(\ref{result=}) is smaller or larger than~1, as illustrated by
Fig.~\ref{fig:distribution}. In the first case the typical scale
$\gamma_n\rightarrow{0}$ as $n\rightarrow\infty$, and the
distribution function for the total~$\Gamma$ is always concentrated
around $\Gamma\sim\eta$ which tends to zero as $\eta\to{0}$. This
perfectly matches the insulating behavior described in
Sec.~\ref{sec:statistics} and shown in Fig.~\ref{figP}b.

Oppositely, for $\Lambda>1$ the scale $\gamma_n\rightarrow\infty$;
we emphasize that the limit $n\to\infty$ should be
taken {\em prior} to $\eta\to{0}$. This means that the
distribution function $P(\Gamma)$ does not shrink to
$\delta(\Gamma)$ as $\eta\to{0}$; in fact, it cannot be determined
from the linearized equation~(\ref{linSCBA=}), the full
self-consistent problem (\ref{SCBArate=}),~(\ref{SCBAspectral=})
should be solved. The divergent linear solution signals the
instability of the insulating state and the onset of the metallic
state.

These arguments enable us to identify the temperature at
which $\Lambda=1$ with the temperature of the metal-insulator
transition:
\begin{equation}
T_c=\frac{\delta_\zeta}{C_2\lambda\ln(1/\lambda)}\,.
\end{equation}

\section{Metal-insulator transition and many-body mobility edge}
\label{sec:MBloc}

In the previous section we considered the decay of a quasiparticle
excitation, and found that possibility or impossibility of such
decay may be viewed as delocalization or localization of this
excitation in the many-electron Fock space. In this section we
discuss macroscopic implications of this picture. How can the very
notion of localization be applied to many-body states?

Consider a many-body eigenstate $|\Psi_k\rangle$ of the
interacting system, with the corresponding eigenenergy~$E_k$. In
the coordinate representation, the many-body wave function
$\Psi_k\left(\left\{\vec{r}_j\right\}_{j=1}^N\right)$ depends on
the coordinates of all $N$ particles in the system.
%The single-particle states forming this many-body state are
%located everywhere in the volume~$\Omega$. Thus, no
%definition in the coordinate space can be constructed.
Let us create an electron-hole pair on top of
$\left|\Psi_k\right\rangle$. The resulting state, which is not an
eigenstate of the system, can be expanded in terms of other
eigenstates:
\begin{equation}\label{cdagPsik=}
\hat{c}^\dagger_\alpha\hat{c}_\beta \left|\Psi_k\right\rangle
 = \sum_{k'}C^{kk'}_{\alpha\beta} \left|\Psi_{k'}\right\rangle;
\quad \sum_{k'}\left|C^{kk'}_{\alpha\beta}\right|^2=1.
\end{equation}
It is possible that the number of terms contributing to the sum is
effectively finite, {\em i.e.}
\begin{equation}
\lim_\mathcal{V\to \infty}
\left[\sum_{k'}\left|C^{kk'}_{\alpha\beta}\right|^4\right]^{-1}
<\infty. \label{eq:2.13b}
\end{equation}
This corresponds to {\em insulating} or {\em localized many-body}
state; excitation can not propagate over all states allowed by the
energy conservation.

The opposite case, when expansion~(\ref{cdagPsik=}) contains an
{\em infinite} number of eigenstates
\begin{equation}
\lim_\mathcal{V\to \infty}
\left[\sum_{k'}\left|C^{kk'}_{\alpha\beta}\right|^4\right]^{-1} =
\infty, \label{eq:2.13c}
\end{equation}
corresponds to {\em metallic} or {\em extended many-body} state.

Developed metallic state is formed when
%electron-electron interaction mixes the excited state with 
expansion~(\ref{cdagPsik=}) involves
all the eigenstates with close enough energies:
\begin{equation}
|C^{kk'}_{\alpha\beta}|^2 \propto ``\delta(E_k
+\omega_{\alpha\beta}-E_{k'})\mbox{''},
\end{equation}
where
$\delta$-function should be understood in the thermodynamic sense:
its width, although sufficiently large to include many states,
vanishes in the limit $\Omega\to\infty$. Only in this regime,
which may also be called {\em ergodic many-body} state, the
electron-electron interaction can bring the system from the
initial Hartree-Fock state to the equilibrium corresponding
to spanning all the states permitted by the energy conservation.
In this case, the averaging over the exact many-body eigenfunction
is equivalent to averaging over the microcanonical distribution,
and temperature~$T$ can be defined as a usual Lagrange multiplier.
It is related to~$E_k$ by the thermodynamic relation:
\begin{equation}
E_k-E_{0}=\int\limits_0^T C_V(T_1)\,d{T}_1, \label{Tdef=}
\end{equation}
where $E_{0}$ is the ground state energy, and
$C_V(T)\propto\Omega$ is the specific heat.

The temperature of the metal-insulator transition, found above, in
fact, determines the extensive {\em many-body mobility edge}
$\mathcal{E}_c\propto\Omega$. In other words, (i)~states with
energies $E_k-E_0>\mathcal{E}_c$ are extended, inelastic
relaxation is possible, and the conductivity
$\sigma_k=\sigma(E_k)$ in this state is finite; (ii)~states with
energies $E_k-E_0<\mathcal{E}_c$ are localized and the
conductivity $\sigma(E_k)=0$.

Let us now assume that the equilibrium
occupation is given by the Gibbs distribution. One could think
that it would still imply the Arrhenius law (\ref{Arrhenius}) for
the conductivity. However, this is not the case for the many-body
mobility threshold. In fact, in the limit $\Omega \to \infty$
\begin{equation}
\sigma(T)=0;\quad T < T_c, \label{transition1}
\end{equation}
where
the critical temperature is determined by Eq.~(\ref{Tdef=}):
%the thermodynamic formula
\begin{equation}
\label{transition2} \int_0^{T_c} d{T}_1\, C_V(T_1)=\mathcal{E}_c.
\end{equation}
 Therefore, the temperature
dependence of the dissipative coefficient in the system shows the
singularity typical for a phase transition.

To prove the relations~(\ref{transition1}),~(\ref{transition2}) we
use the Gibbs distribution and find
\[
\sigma(T)=\sum_k P_k \sigma(E_k) =\frac{\int_0^\infty{d}E\,
e^{S(E)-E/T}\sigma(E)} {\int_0^\infty dE\, e^{S(E)-E/T}},
\]
where the entropy $S(E)$ is proportional to volume, and $E$ is
counted from the ground state. The integral is calculated in the
saddle point or in the steepest decent approximations, exact for
$\Omega\to\infty$. The saddle point $E(T)$ is given by
\[
\left.\frac{dS}{dE}\right|_{E=E(T)}=\frac{1}{T}\,.
\]
Taking into account $\sigma(E)=0$ for $E < \mathcal{E}_c$ we find
\begin{eqnarray*}
&&\sigma(T)=\sigma\left[E(T)\right], \quad E(T)>\mathcal{E}_c;
\\
&& \sigma(T) \propto
\exp\left(-\frac{\mathcal{E}_c-E(T)}{T}\right); \quad E(T)<
\mathcal{E}_c
\end{eqnarray*}
As both energies entering the exponential are extensive,
$E(T),\mathcal{E}_c \propto \Omega$, we obtain
(\ref{transition1}),~(\ref{transition2}).

To be able to establish the thermal equilibrium in such insulating
state the system should be coupled to an external bath
({\em i.e.}, phonons). The presence of the finite electron-phonon
interaction  (as phonons are usually delocalized), smears out the
transition, and $\sigma(T)$ becomes finite for any
temperature. Nevertheless, if electron-phonon interaction is weak, the
phenomenon of the many-body metal-insulator transition manifests
itself as a sharp crossover from phonon induced hopping at $T<T_c$ to
the conductivity independent of the electron-phonon coupling at
$T>T_c$.

\section{Conclusions and perspectives}\label{sec:conclusions}

\begin{figure}
\includegraphics[width=0.7\textwidth]{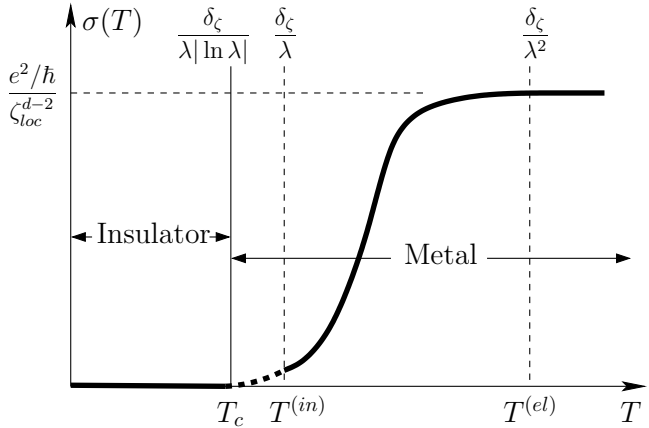}
\caption{Schematic temperature dependence of the dc conductivity
  $\sigma(T)$. Below the point of the many-body metal-insulator
  transition, $T<T_c$, no inelastic relaxation occurs and
  $\sigma(T)=0$. Temperature interval $T\gg{T}^{(in)}>T_c$ corresponds
  to the developed metallic phase, where Eq.~(\ref{developedmetal=})
  is valid. At $T\gg{T}^{(el)}$ the high-temperature metallic
  perturbation theory (Altshuler and Aronov, 1985) is
  valid, and the conductivity is given by the Drude formula.
}\label{fig:summary}
\end{figure}

We have considered the inelastic relaxation and transport at low
temperatures in disordered conductors where all single-particle states
are localized, and no coupling to phonons or any other thermal bath is
present. The main question is whether electron-electron interaction
alone is sufficient to cause transitions between the localized states
producing thereby a finite conductivity. The answer to this question
turns out to be determined by the Anderson localization-delocalization
physics in the many-particle Fock space, and is summarized on
Fig.~\ref{fig:summary}.

It should be emphasized that the many-body localization, which we discuss 
in this paper, is qualitatively different from conventional finite 
temperature Metal to Insulator transitions, such as formation of a 
band insulator due to the structural phase transition or Mott-Hubbard 
transition. In these two cases, at a certain
temperature $T^*$ a gap appears in the spectrum of charge excitation
(Mott insulator) or all excitations (band insulator). In both cases
the conductivity remains finite although exponentially small as long
as $T>0$. This is not the case for many-body localization, which
implies exactly zero conductivity in the low-temperature phase.

Is the many-body localization a true thermodynamical phase transition
with corresponding singularities in all equilibrium properties?
This question definitely requires
additional studies, however, some speculations can be put forward.
The physics described in the present paper is associated with
the change of the characteristics of the many-body wavefunctions.
It is well known that for non-interacting systems
localization-delocalization transition does not affect the average
density of states, {\em i.e.}, it does not manifest itself in
any macroscopic thermodynamic properties. Application of the same
logic to the exact many-body eigenvalues would indicate that the
many-body localization transition
is not followed by any singularities in the static specific heat, etc.
On the other hand, at this point we can not rule out the possibility
that this conclusion is an artifact of treating the real parts of the
electron self-energies
with an insufficient accuracy. Most likely scenario, to our opinion, 
is that the insulating phase 
behaves like a glass (spin or structural) and demonstrates all the glassy 
properties (Fisher and Hertz, 1991; Bouchaud {\em et al.}, 1998), like
absence of ergodicity (even when some 
coupling with  phonons is included), effects of aging, etc. 
Discussion of the equilibrium susceptibilities in the latter case
becomes quite meaningless. 

The quantitative theory built in BAA~paper assumes that the
interaction is weak. On the other hand, qualitative consideration
of the localization of many-body excitations does not rely on this
assumption. The important ingredients are (i)~localization of
single-particle excitations, and (ii)~Fermi statistics. 
Consider, as an example, Wigner crystal (Wigner, 1934).
It is well known that strong enough interaction leads to a spontaneous
breaking of the translational symmetry in $d$-dimensional clean
systems at $d\geq{2}$. In a clean system Wigner crystallization is either
a first-order phase transition ($d=3$), or a Kosterlitz-Thouless
transition ($d=2$). Even weak disorder destroys both translational and
orientational order (Larkin, 1970) and pins the crystal. The symmetry
of this state is thus not different from the symmetry of a liquid, and
the thermodynamic phase transition is commonly believed to be reduced
to a crossover.

We argue that the many-body localization provides the correct
scenario for the finite-temperature ``melting'' transition between the
insulating phase, which may be called ``solid'', and the metallic
phase, which may be called ``liquid''. Indeed, the conductivity of the
pinned Wigner crystal is provided by the motion of defects. At low
temperatures and in the absence of the external bath, all defects are
localized by the one-particle Anderson mechanism. Phonon modes of the
Wigner crystal are localized as well, so the system should behave as a
many-body insulator. As the temperature is increased, the many-body
metal-insulator transition occurs. It is not clear at present,
whether it occurs before or after the crystalline order is destroyed
at distances smaller than Larkin's scale. Construction of effective
theory of such a transition is a problem which deserves further
investigation.

%\bibliographystyle{OUPnamed}
%\bibliography{Keldysh75}
\end{document}